# An Improved Quantum Algorithm of the Multislice Method

Y.C. Wang[1], Y. Sun[3#], Z.J. Ding[1,2*]

[1]*Department of Physics, University of Science and Technology of China, Hefei, Anhui 230026, People's Republic of China*

[2]*Hefei National Research Center for Physical Sciences at the Microscale, University of Science and Technology of China, Hefei, Anhui 230026, People's Republic of China*

[3]*Department of Physics, Xiamen University, Xiamen, Fujian 361005, People's Republic of China*

[#]e-mail: yangsun@xmu.edu.cn

[*]e-mail: zjding@ustc.edu.cn

The multislice method is an important algorithm for electron diffraction and image simulations in transmission electron microscopy. We have proposed a quantum algorithm of the multislice method based on quantum circuit model previously. In this work we have developed an improved quantum algorithm. We reconstruct the phase-shifting quantum circuit without using the multi-controlled quantum gates, thereby significantly improve the computation efficiency. The new quantum circuit also allows further gate count reduction at the cost of a controllable error. We have simulated the quantum circuit on a classical supercomputer and analyzed the result to prove the feasibility and correctness of the improved quantum algorithm. We also provide proper parameter settings through testing, allowing the minimization of the necessary number of quantum gates while limiting the relative error within 1%. This work demonstrates the potential of applying quantum computing to electron diffraction simulations and achieving quantum advantages.

**Keywords:** Quantum algorithm, Multislice method, Electron scattering and diffraction, Quantum computing, Quantum simulation

# 1. Introduction

With various advancements in the physical realization of quantum computers [1-3], quantum computing has received tremendous attention in recent years. By utilizing quantum entanglement and quantum superposition, quantum computing is able to efficiently solve certain complex problems that are intractable for classical computers. For example, Shor's algorithm [4] can solve the integer factorization problem in a polynomial time. And Grover's algorithm [5] can substantially speed up the search in unstructured databases. Quantum simulation [6] is a major application of quantum computing, with the idea, proposed by Feynman, of using a quantum system as a quantum computer to simulate another quantum system efficiently [7,8].

In our previous work, we have proposed a quantum algorithm of the multislice method based on quantum circuit model to simulate electron scattering and diffraction using quantum computing [9]. The multislice method is a widely used method in electron microscopy to simulate the propagation of the high-energy electron beam in the three-dimensional atomic potential field. The original multislice method was proposed by Cowley [10-12] and has been improved and extended many times since [13-18]. Nowadays the multislice method has applications in various areas of electron beam related techniques, including simulations of TEM images [19,20], STEM images [21-23], Bohmian quantum trajectories [24,25], electron energy loss spectra (EELS) [26,27]. Unlike Bloch wave method [28,29], the other commonly used method for simulating electron diffraction, the multislice method can in principle handle complex nonperiodic potential fields. But as object size in simulation increases, the computation cost of the classical multislice method increases exponentially due to the repeatedly use of the fast Fourier transform (FFT) [30]. In the previous quantum algorithm, we replaced FFT with the quantum Fourier transform (QFT) [31,32] and constructed phase-shifting quantum circuits that work together to perform iterations, achieving a preliminary quantum speedup. One potential problem was that we used a large number of multi-controlled quantum gates in phase-shifting circuits. Current quantum processors usually have fixed topology and will have to compile multi-controlled gates into one- and two-qubit gates, which may require linear cost related to the number of control qubits [33]. This additional factor of cost may nullify the quantum advantage of the quantum algorithm on practical quantum hardware.

In this paper, we reconstruct phase-shifting circuits with only one- and two-qubit quantum gates, which are comparable in number to the multi-controlled gates used before. Our new phase-shifting quantum circuit is based on the quantum circuit for diagonal unitary given by Welch et al. [34], who gave a one-dimensional example of Eckart barrier in their work. We extend their approach to two-dimensional situations and apply it to the multislice method to deal with three-dimensional potential field. Then we introduce a truncation approach that allows cutting off some unimportant terms after Walsh transform to further reduce the quantum gate number. The truncation thresholds are carefully set according to an empirical formula derived from tests in order to achieve a trade-off between the quantum gate number and the truncation error. In this way, the improved quantum algorithm of the multislice method can use a significantly smaller number of one- and two-qubit quantum gates to perform the same simulations as before with a sufficiently small additional error.

## 2. Quantum multislice algorithm

The basic concept of the multislice method is to divide the sample into a series of equally spaced slices perpendicular to the electron beam. When slices are thin enough, the electron wave function can be considered to propagate freely within a slice, and only interacts with a two-dimensional potential field projected on each interface. More specifically, considering one iteration of the electron wave function $\psi(\mathbf{r})$ from the slice $t$ to the slice $t+1$, we have

$$\psi_{t+1}(\mathbf{r}) = \exp\left(-i\pi d\lambda |\mathbf{Q}|^2\right) \exp\left(i\sigma d V_t(\mathbf{r})\right) \psi_t(\mathbf{r}), \tag{1}$$

where $\sigma = \dfrac{me\lambda}{2\pi\hbar^2}$. $\lambda$ is the electron wavelength, $m$ the mass of an electron, $e$ the elementary charge, $d$ the thickness of a slice. $\psi_t(\mathbf{r})$ and $V_t(\mathbf{r})$ are the wave function and the potential energy at slice $t$, $\boldsymbol{r}$ represents the space coordinates $(x, y)$ in the $xy$-plane, $\boldsymbol{Q}$ represents the space frequency $(Q_x, Q_y)$ in the plane. Ishizuka et al. proposed that the maximum slice thickness $d$ should be $\sim kd_P$ in order to get a stable result close to the limiting value, where $k$ is the wavenumber of the incident electrons and $d_P$ is the distance over which the potential does not change by an appreciable fraction [14].

In high-energy approximation, the effect of the potential to the electron momentum can be considered as a small perturbation, hence, the kinetic energy of the electron, $E_k$, is always equal to the initial kinetic energy, $eU$, where $U$ is the accelerating voltage of the electron beam. Considering the relativistic effect, the relativistic mass of electron can be given by

$$m = m_0 + eU/c^2, \tag{2}$$

and the electron wavelength can be given by

$$\lambda = \frac{hc}{\sqrt{2E_0 E_k + E_k^2}}, \tag{3}$$

where $E_0 = m_0 c$, $m_0$ is the rest mass of electron, $c$ is the speed of light in vacuum.

The potential energy term $\exp(i\sigma \Delta z V_t(\boldsymbol{r}))$ is diagonal in the coordinate representation, and the kinetic energy term, i.e. the propagator, $\exp(-i\pi d\lambda |\boldsymbol{Q}|^2)$ is diagonal in the momentum representation. We use Fourier transform (FT), denoted by $F$, and inverse Fourier transform (iFT), denoted by $F^{-1}$, to transform the wave function between two representations. In this way, one iteration can be written as a unitary operator $F^{-1}PFV$, where both potential operator $V$ and kinetic operator $P$ are diagonal. Fig. 1 show the process of one iteration.

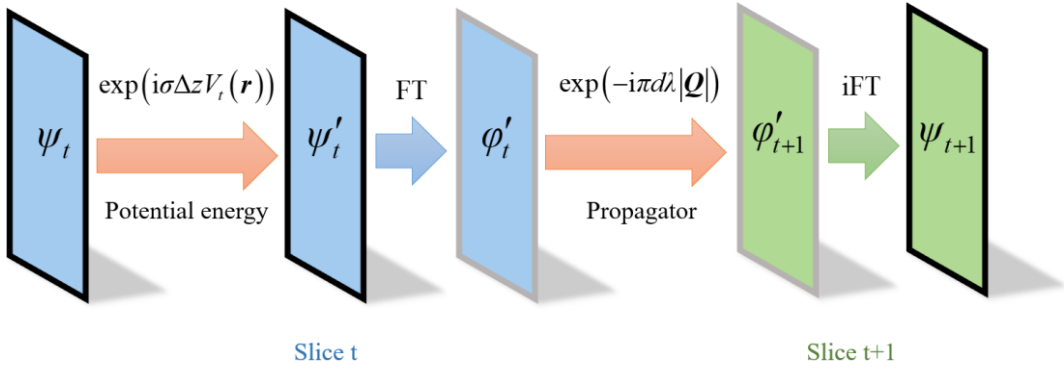

Fig. 1. The flowchart of the wave function iteration between two slices in the multislice method. The black frame represents the coordinate representation, while the gray one represents the momentum representation.

In the quantum algorithm of the multislice method, qubits and quantum gates form a quantum circuit. Input information is encoded into the quantum state of qubits and processed by quantum gates to achieve calculation. We use two-dimensional amplitude encoding to store a slice of the wave function. The two-dimensional wave function is discretized into a $N \times N$ complex matrix where $N = 2^n$. Each matrix element represents the amplitude of the wave function at a coordinate in the real space. To store this matrix, we use a quantum register with $2n$ qubits, which has $2^{2n}$ basis states and stores each matrix element by the amplitude of a basis state. More specifically, we divide the quantum register into two parts, each with $n$ qubits. According to the binary code, the lower $n$ qubits represent the *x*-coordinate and the higher $n$ qubits represent the *y*-coordinate. In this way, each basis state corresponds to a coordinate point. Then we can use the quantum Fourier transform (QFT) to achieve the Fourier transform in the iteration, which is exponentially faster than the fast Fourier transform (FFT) that is typically used in the classical multislice method. The quantum circuit for one-dimensional QFT is as Fig. 2. Here $H$ represents the Hadamard gate, and $R_m$ represents a controlled phase-shifting gate corresponding to the transformation matrix

$$R_m = \begin{bmatrix} 1 & 0 \\ 0 & e^{2\pi i/2^m} \end{bmatrix}. \tag{4}$$

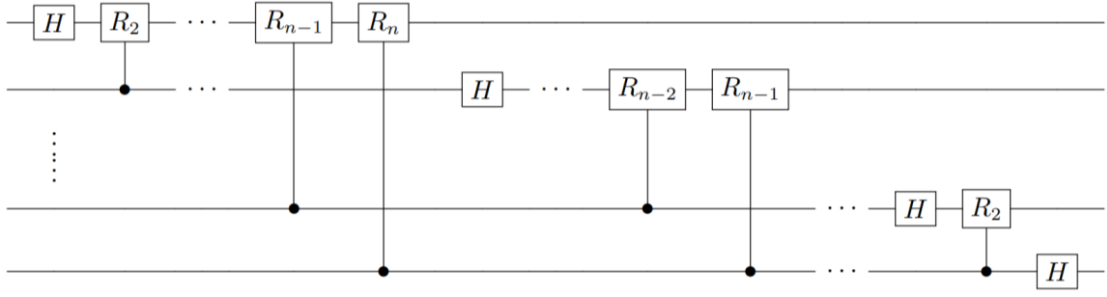

Fig. 2. The quantum circuit for 1D QFT.

The circuit can be run in reverse to perform the inverse quantum Fourier transform (iQFT). It can be proven that we can perform a two-dimensional FT on the quantum state matrix by applying two *n*-qubit iQFT to both lower *n* qubits and higher *n* qubits in parallel [9].

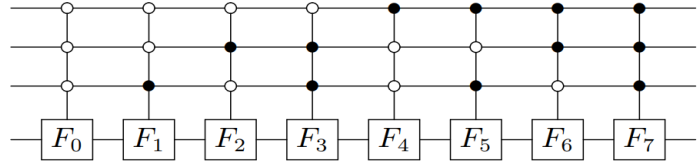

Fig. 3. The previous version of the phase-shifting quantum circuit for 4 qubits, *n*=2.

Apart from QFT circuits and iQFT circuits, there are phase-shifting quantum circuits that apply diagonal phase-shifting operator on the quantum state, satisfying:

$$U|x\rangle = e^{if(x)}|x\rangle. \tag{5}$$

In the previous work, we used a simple phase-shifting circuit shown in Fig. 3 to implement this operator, where $k \in \{0,1,\cdots 2^{2n-1}-1\}$, and $F_k$ corresponds to the matrix

$$F_k = \begin{bmatrix} e^{if(2k)} & 0 \\ 0 & e^{if(2k+1)} \end{bmatrix}. \tag{6}$$

Then the whole quantum circuit for one iteration is as shown Fig. 4, where $V$ and $P$ represent respectively the phase-shifting quantum circuits for the potential operator and the free propagation operator. By preparing an initial state on qubits and applying this circuit iteratively, we can simulate the evolution of the electron wave function. We use the most basic input state, plane wave, as an example. To prepare this initial state, we only need an all-zero state and then apply a Hadamard gate on each qubit to get a uniform superposition state. Other arbitrary forms of initial quantum states can be obtained by a sublinear quantum circuit with complexity no more than $O(N)$ [35], and there is a more efficient method for a sparse state [36]. After a certain number of iterations, the position distribution and momentum distribution of the electron can be extracted from the output quantum state.

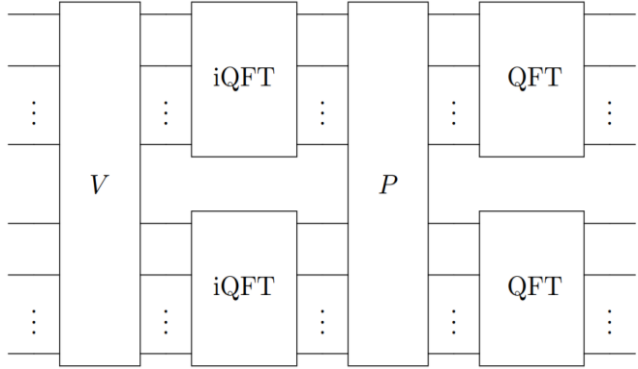

Fig. 4. Diagram of the quantum circuit required for the evolution of one slice of electron wave function.

Notice that the phase-shifting quantum circuit utilizes a large number of multi-controlled quantum gates as we can see in Fig. 3, which may be inefficient for practical near-term quantum hardware [33]. In the next section, we will focus on the optimization of the phase-shifting circuit for addressing this problem.

In practical quantum computing, how to efficiently extract information from the quantum state is also a problem worthy of study. A basic approach is to perform a standard projective measurement after each run and to repeat the quantum simulation. Each run provides a single sample, and the measured results will form a distribution which matches with the probability distribution of the electron state. Here the number of repetitions needed depends on the shape of the distribution and the statistical error we can accept. A sampling demonstration is presented in our previous work [9], where we reconstruct the electron diffraction pattern by using up to 5000 samples in the momentum representation. In this case, the required sample number is much smaller than the number of grid points and does not increase with $N$, adding only a constant factor to the complexity, because in the momentum representation the samples have a high probability of landing on the fixed diffraction peaks and forming the pattern easily. Advanced extraction of full quantum state would involve quantum state tomography [37] which will not be discussed in this paper. Besides extracting information directly from the output quantum state, one may use the output quantum state as input for further possible quantum algorithms to bypass the full extraction. One may also use it for a fast sampling of electron probability distribution as a sub-algorithm of other simulations.

### 3. Optimization of the phase-shifting circuit

We reconstruct the phase-shifting quantum circuit by using the Walsh transform [38]. The Walsh transform expands a discrete function by a series of orthogonal basis functions named Walsh functions. For a 1D function $f(x)$ where $x \in \{0, 1, \cdots, N-1\}$, $N = 2^n$, the Walsh functions in Hadamard order are defined as follow:

$$T_u(x) = (-1)^{\sum_{i=1}^{n-1} x_i u_i}, \tag{7}$$

where $u \in \{0, 1, \cdots, N-1\}$, and $x_n$ represents the $n$th bit in the binary expansion of $x$, i.e., $x = (x_{n-1} x_{n-2} \cdots x_0)_2$, $u = (u_{n-1} u_{n-2} \cdots u_0)_2$. Under this definition, the transform is named Walsh-Hadamard transform, which can be written as follow:

$$W(u) = \frac{1}{N} \sum_{x=0}^{N-1} f(x) (-1)^{\sum_{i=0}^{n-1} x_i u_i}, \tag{8}$$

And $f(x)$ can be represented as a linear combination of Walsh basis functions, called Walsh expansion:

$$f(x) = \sum_{u=0}^{N-1} W(u) T_u(x). \tag{9}$$

For example, the 1D Walsh basis functions when $N = 8$ are shown in Fig. 5.

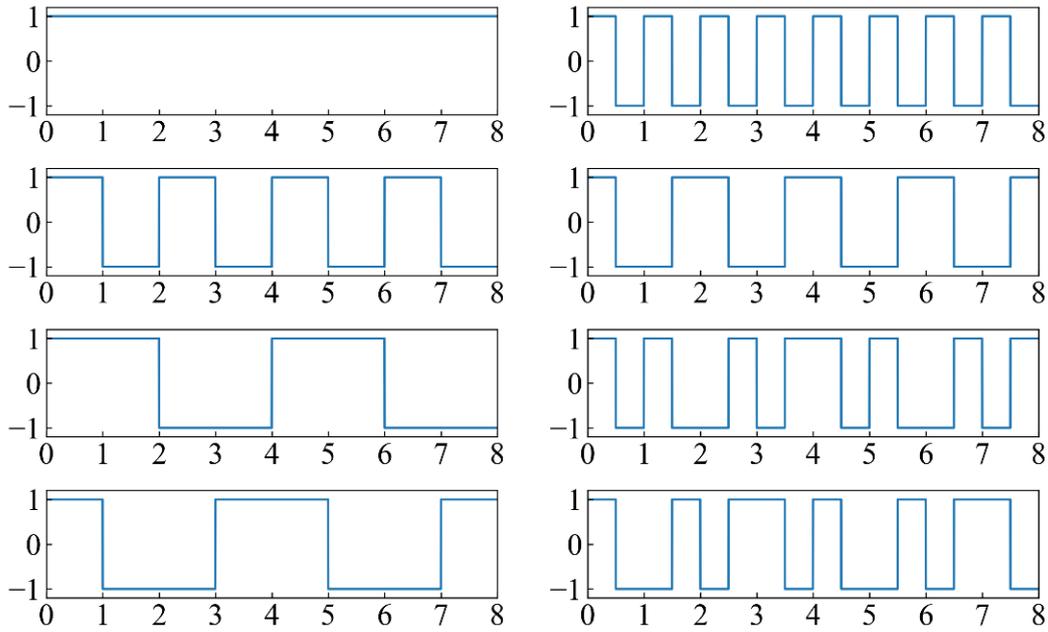

Fig. 5. The 1D 8-point Walsh functions in Hadamard order.

For 2D situation, we just need to combine the binary expansions of $x$ and $y$ together, and apply the 1D formula. For example, the 2D Walsh basis function when $N = 4$ are shown in Fig. 6. Similar to the 1D situation, an arbitrary $4\times 4$ discrete function can be seen as a linear superposition of these 16 basis functions.

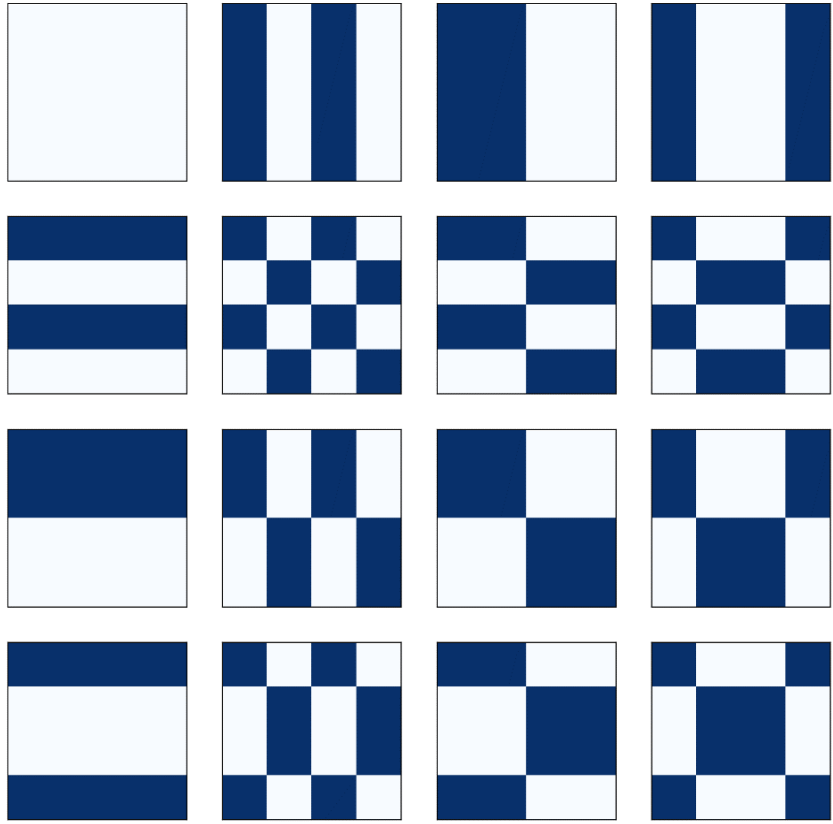

Fig. 6. The 2D $4\times4$-point Walsh functions. The white pixel represents 1 while the blue pixel represents -1.

Interestingly, the patterns of Walsh functions are well suited for a quantum circuit. In our previous version of the phase-shifting quantum circuit, we were actually manipulating the phase of each basis state individually, which is hard because of the superposition. Multiple control qubits must be attached to the phase-shifting quantum gate as shown in Fig. 3 to shift the phase of a single basis state, while a phase-shifting gate without control qubits will result in a phase shift for half of the basis states, forming a pattern that interestingly just matches a Walsh function. We can then construct a phase-shifting quantum circuit for the phase-shifting operator $|x\rangle \to e^{if(x)}|x\rangle$ according to the Walsh transform of $f(x)$. An 1D example has been given by Welch et al. [34], and we extend their approach to two-dimensional situations.

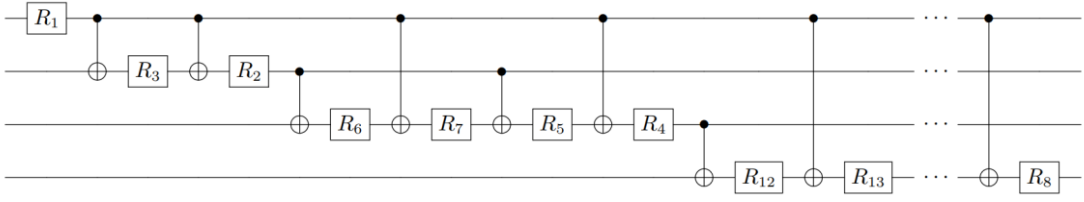

Fig. 7. The improved version of the phase-shifting quantum circuit for 4 qubits, *n*=2.

The improved phase-shifting quantum circuit for 4 qubits is shown in Fig. 7 as an example, consisting of $N^2-1$ phase-shifting quantum gates and a number of CNOT quantum gates between them, with no multi-controlled gate used. Each phase-shifting gate corresponds to one term of the Walsh expansion of the phase-shifting operator, i.e. a scaled Walsh function with a certain index. The amount of the phase shift is determined by each coefficient of this Walsh function. The first Walsh function with index 0 corresponds to a global phase shift, which is generally considered to be meaningless in quantum physics. These phase-shifting gate are ordered according to the Gray code [39] to minimize the number of CNOT gates required. Gray code is a binary code system where two successive values differ in only one bit. CNOT gates between two phase-shifting gates are determined by the difference in the binary expansions of the two neighboring indexes, i.e. the Hamming distance [40]. In this order, only one CNOT gate is required between every two phase-shifting gates.

This phase-shifting quantum circuit can be generalized to two dimensions. In our algorithm, the wave function of a slice is stored as a $N \times N$ matrix with two indexes $x$ and $y$. We can concatenate the binary codes of two indexes together and decode it as a new index $r$, $r \in \{0, 1, \cdots N^2-1\}$, which satisfies:

$$r = Ny + x. \tag{10}$$

In this way, we reshape a 2D matrix into a 1D array while the correspondence between the phase shifts that need to be applied to each element of the wave function remains. Specifically, to apply a 2D phase-shifting operator satisfying:

$$U|x, y\rangle = e^{if(x,y)}|x, y\rangle, \tag{11}$$

we set a $g(r)$ satisfying:

$$g(r) = f(x, y), \tag{12}$$

where $x = r \mod N$, $y = r - Nx$. Then applying $U$ to the 2D matrix of the wave function is equivalent to applying $U'$ to the reshaped 1D array, where $U'$ satisfies:

$$U'|r\rangle = e^{ig(r)}|r\rangle. \tag{13}$$

We can represent $U'$ as a $N^2 \times N^2$ diagonal matrix:

$$U' = \text{diag}(e^{if(0,0)}, e^{if(1,0)}, \cdots, e^{if(N,0)}, e^{if(0,1)}, \cdots, e^{if(N,1)}, \cdots, e^{if(N,N)}). \tag{14}$$

The quantum state in qubits need no change for the reshaping, it can be naturally decoded in two ways because of our encoding method where the lower $n$ qubits represent the *x*-coordinate and the higher $n$ qubits represent the *y*-coordinate.

We also need to apply Walsh transform to the potential operator matrices and the kinetic operator matrix as a classical preprocessing step to determine the parameters in the phase-shifting quantum circuits. The complexity of the fast Walsh-Hadamard transform (FWHT) is $O(N \log N)$, which is the same as the complexity of the FFT. Nevertheless, this preprocessing step is efficient and will not become a new bottleneck of the whole algorithm. First, unlike the FFT, which uses a large number of multiplications, the FWHT uses only additions, which are much faster for classical computers. What is more, the results obtained from the Walsh transform are reusable. For the potential field in the lattices, we only need to compute the Walsh transform of the potential field slices within one lattice. The computational cost does not increase continually with the number of lattice layers. It can also be reused for different input electron states. The kinetic term is even simpler and remains constant, requiring only one Walsh transform to be able to reuse throughout the entire simulation process.

In summary, our improved quantum algorithm still follows the diagram in Fig. 4, using the improved phase-shifting quantum sub-circuit for $2n$ qubits to achieve the phase-shifting operations $V$ and $P$. The parameters of the quantum circuit are determined by the WHT preprocessing steps for operators $V$ and $P$, which run on a classical computer.

So far, the quantum algorithm has been an exact quantum version of the classical multislice method. However, if we allow for some additional error, we can further reduce the number of quantum gates required. After the Walsh transform of the operator, we set a truncation threshold and cut off the terms with coefficients that are less than the threshold. In this way, we can ignore and skip those Walsh basis functions with low contributions and achieve an approximation of the operator using a quantum circuit with lower depth. It is worth mentioning that this optimization can only be done in the quantum algorithm, not in the classical algorithm. In the classical algorithm, without the quantum superposition, operating half of the phases to execute a phase shift that matches a Walsh function pattern is inefficient. It is also not easy to separate the components of each Walsh basis function in the operator and skip some of them.

In this paper, we use a relative truncation threshold $\tau$ and cut off a term in Walsh decomposition if its coefficient $w$ is lower than $\tau w_{\max}$ where $w_{\max}$ represents the maximum coefficient in the Walsh decomposition and $\tau \in [0,1)$. As $\tau$ increases, the number of required quantum gates will decrease, while the truncation error will increase. We test the quantum algorithm with different qubit number and truncation threshold settings and obtain an empirical formula to appropriately set the truncation threshold separately for the potential energy terms and the kinetic energy terms, ensuring that the truncation error is kept within a certain range while minimizing the number of required quantum gates. Detailed truncation threshold settings and error analysis will be discussed together with the feasibility verification test in the next section.

## 4. Results and discussion

Current quantum hardware does not yet have enough qubits and sufficient fidelity to run this quantum algorithm, so we use a classical supercomputer to simulate a quantum circuit with a Python package "pyqpanda" [41,42] at the level of qubit and quantum gate, and run the quantum algorithm on the virtual quantum circuit. We simulate the same scenario as in the previous work, where electrons at different energies incident into a thick Au specimen, to facilitate a comparison between the previous quantum algorithm and the improved one.

The following formula is used to calculate the atomic potential field in the crystal [43]:

$$\varphi(r) = \frac{h^2}{2\pi m_0 e} \sum_i a_i \left(\frac{4\pi}{b_i + B}\right)^{3/2} \exp\left(-\frac{4\pi^2 r^2}{b_i + B}\right), \quad (15)$$

where $a_i$ and $b_i$ $(i=1,2,3,4)$ are fitting parameters, $B$ is the Debye-Waller factor that depends on temperature.

To verify the correctness of the improved quantum algorithm of the multislice method, we have simulated electron diffraction using the same parameters with three different algorithms: the classical multislice algorithm, the previous quantum algorithm, and the present improved one, obtaining identical results. Here we keep all the terms in the improved phase-shifting part to avoid any truncation error. The calculation results can be extracted from different perspectives, and here we choose to display the side cross section of the electron probability density distribution for easy comparison. The position of the cross section is shown in Fig. 8(a), and the simulation results are shown in Fig. 8(b)-8(d). This indicates that the improved phase-shifting quantum circuit can accurately replace the previous version, thereby avoiding the use of multi-controlled quantum gates.

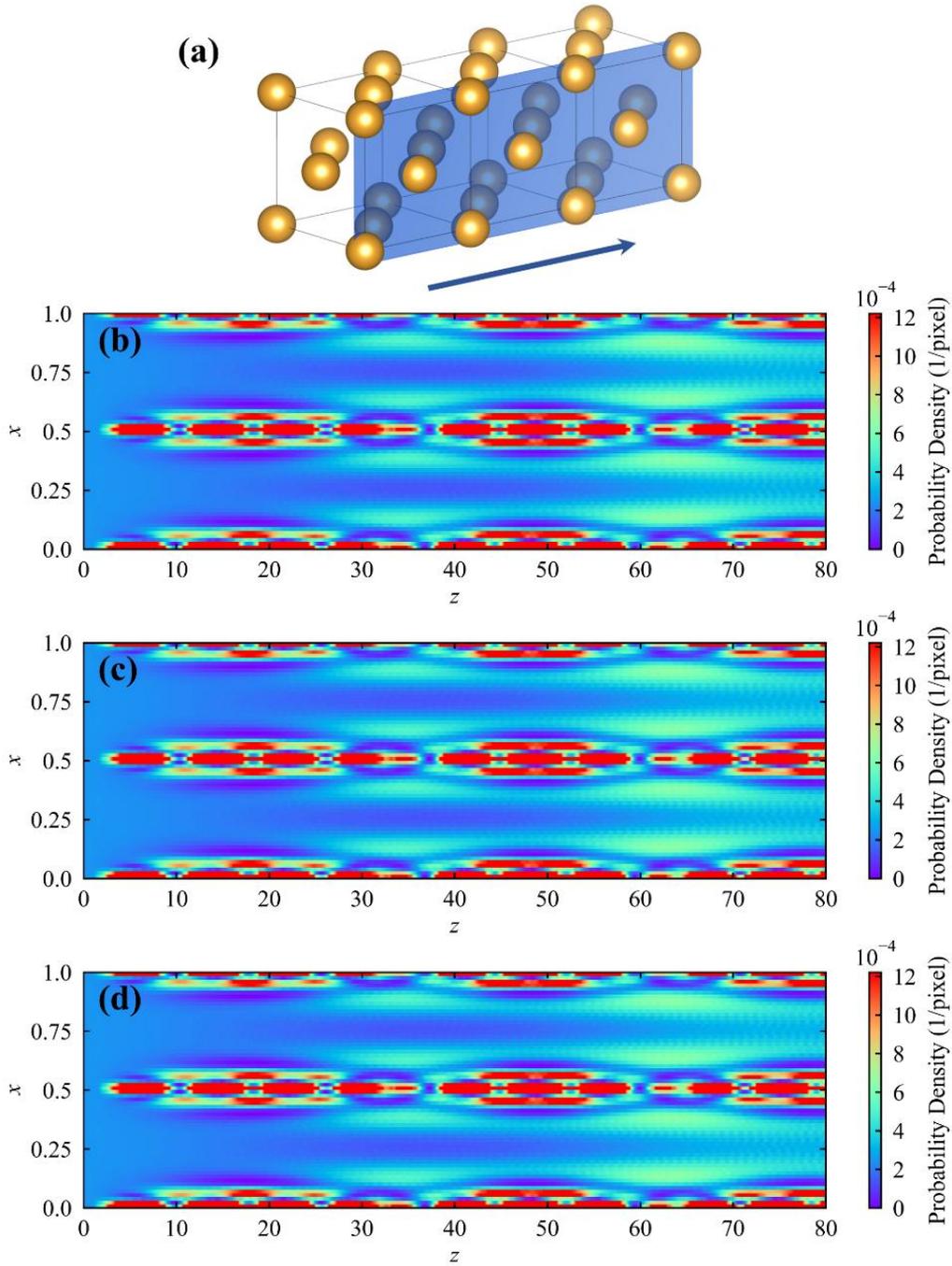

Fig. 8. (a) The diagram of a side cross section atomic plane parallel to the electron beam direction. The arrow indicates the incident direction of the electron beam. (b-d) The simulation results of three algorithms for plane wave incident electrons at 100 keV: (b) the previous quantum algorithm, (c) the improved quantum algorithm, (d) the classical algorithm. We use lattice constant as the unit length. Each unit cell is divided by 16 slices in $z$-direction, satisfying the limitation of the maximum slice thickness. $n = 6$, meaning 12 qubits are used in the quantum algorithm.

Next, we test different truncation thresholds based on this set of parameters. As described in the previous section, we apply the Walsh-Hadamard transform to the potential operators and the kinetic operator, then we cut off the terms of the Walsh decompositions with coefficients that are less than $\tau w_{max}$.

Fig. 9 visually shows the effect of Walsh transform and truncation on the operator matrices. There are 16 different potential operators for the 16 slices per cell, and we choose the one on the atomic plane. Meanwhile, the kinetic operator is the same for every slice. It is clear from Fig. 9 that as the truncation threshold increases, the approximation of the operator becomes increasingly distorted because more terms are cut off.

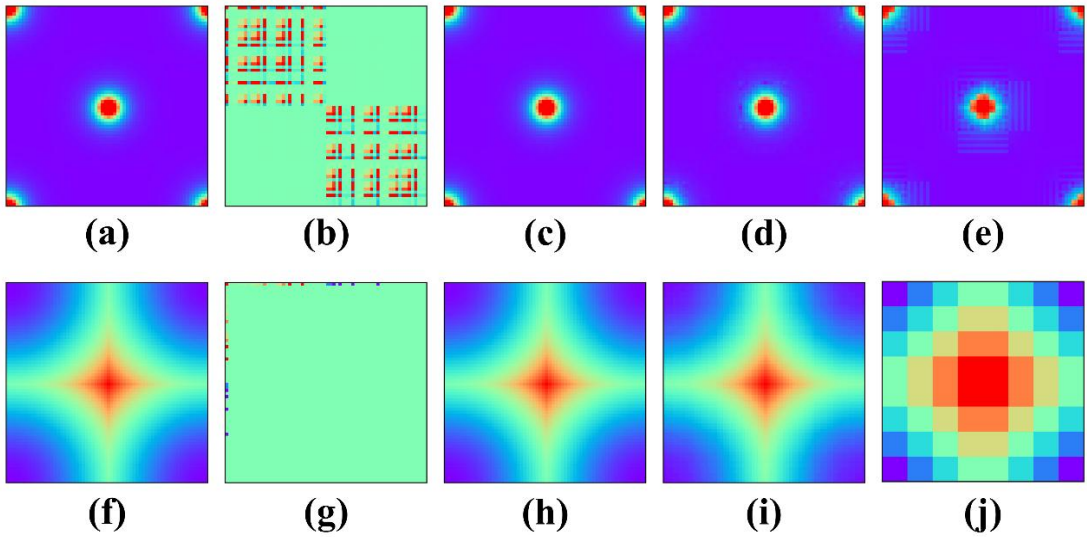

Fig. 9. (a) The potential operator. (b) The potential operator after Walsh transform. (c)-(e) The approximate potential operators when $\tau$ are (c) 0.001, (d) 0.01, (e) 0.1. (f) The kinetic operator. (g) The kinetic operator after Walsh transform. (h)-(j) The approximate kinetic operators when $\tau$ are (h) 0.001, (i) 0.01, (j) 0.1.

Fig. 10 shows the effect of the truncation on the simulation results. Similarly, we use the side cross section of the electron probability density distribution to illustrate. It can be seen that as the truncation threshold increases, the simulation results also become increasingly distorted, being consistent with Fig. 9.

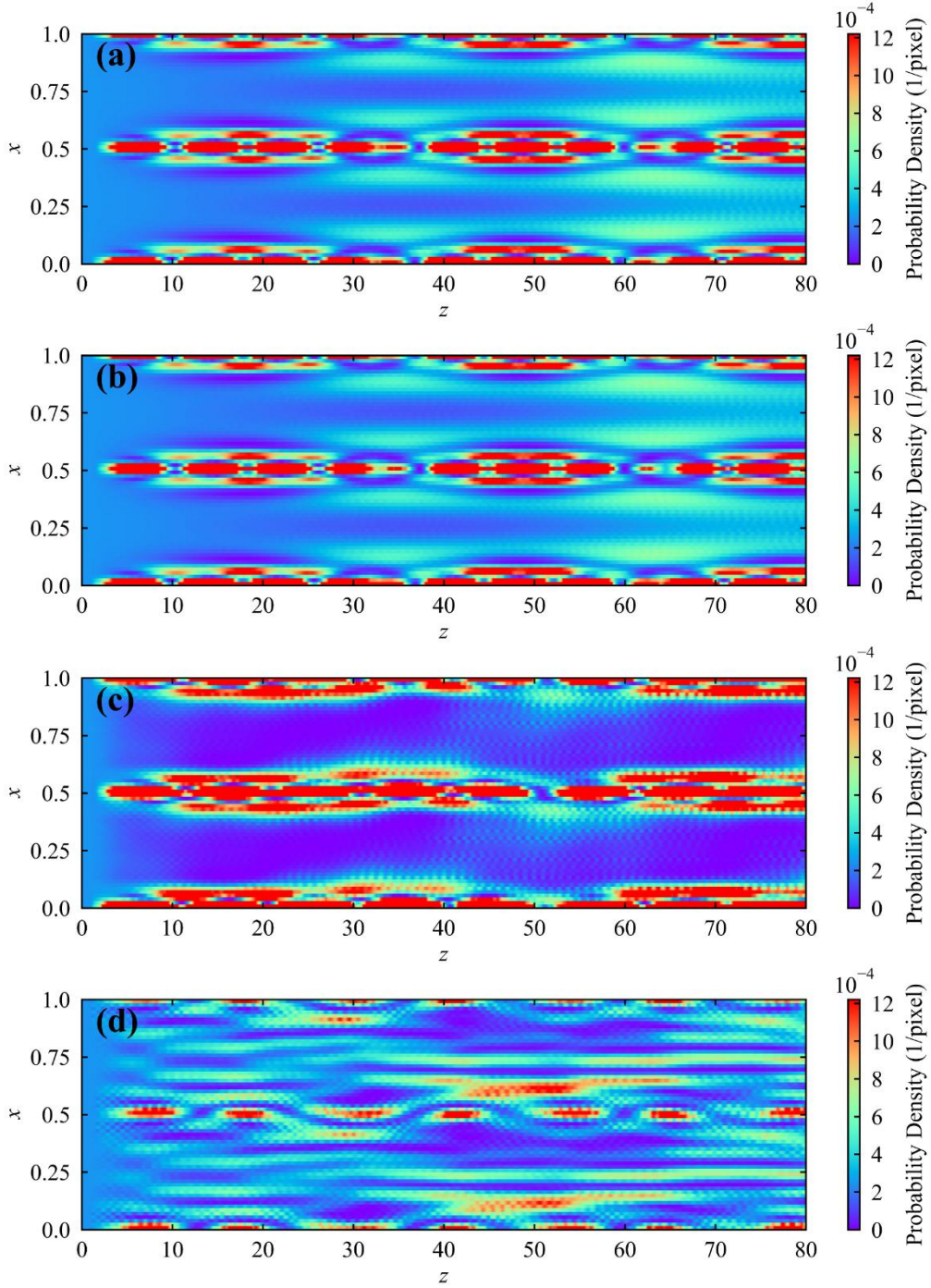

Fig. 10. The simulation results of the electron probability density distribution using the same truncation threshold $\tau$ for the kinetic operator and the potential operator: (a) $\tau = 0$ (exact); (b) $\tau = 0.001$; (c) $\tau = 0.01$; (d) $\tau = 0.1$.

It is worth noting that the distribution patterns of the Walsh coefficients for the potential operator and the kinetic operator differ significantly as shown in Fig. 9. For the potential operator, about half of the coefficients are close to zero because of its symmetry, as shown in green. The remaining coefficients have a relatively scattered distribution in terms of their magnitudes. As for the kinetic operator, due to its simpler analytical expression, almost all coefficients are zero, with only a few non-zero coefficients distributed along the first row and first column. This indicates that much fewer Walsh basis functions are required to approximate the kinetic operator than the potential operator.

To further explore suitable truncation thresholds for different operators, we then cut off potential terms or kinetic terms separately with different truncation thresholds and analysis truncation errors quantitatively. We use the average relative error defined as follow to quantify the truncation error:

$$\varepsilon = \frac{\sum |x - \hat{x}|}{\sum |\hat{x}|}, \tag{16}$$

where $x$ represents the approximate value and $\hat{x}$ represents the exact value.

We start with an initial screening where all zero terms are removed, then we apply further truncation separately to either kinetic or potential terms. During the testing, we first keep all the non-zero kinetic terms while gradually increasing the truncation threshold $\tau_V$ for potential terms and denote the total number of the remaining terms as $s_V$. Similarly, we keep all the non-zero potential terms while gradually increasing the truncation threshold $\tau_P$ for kinetic terms and denote the total number of the remaining terms as $s_P$. The results are shown in Fig. 11. The number of quantum gates in the quantum circuit is proportional to the number of the remaining terms, so fewer terms mean higher efficiency. We can see that as $\tau_V$ increases, the truncation error increases and the term number decreases, both changing linearly and steadily in log scale, making it very suitable for the trade-off between the two. However, as $\tau_P$ increases, the term number only slightly decreases, while the truncation error rises sharply to an unacceptable level after a certain point. This is because that the coefficients of the potential terms are distributed more evenly and widely, while most coefficients of the kinetic terms are zero and the remaining few Walsh terms contribute so significantly that they cannot be ignored.

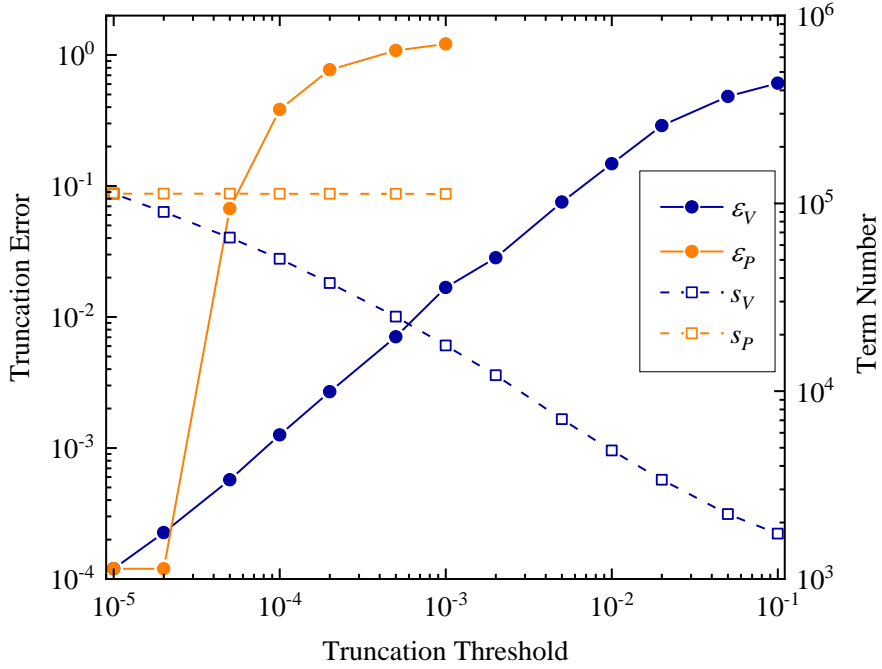

Fig. 11. The truncation errors and the number of the total remaining terms when a gradually increasing truncation threshold is applied to either potential or kinetic terms. $\varepsilon_V$ and $\varepsilon_P$ represent the truncation errors for potential and kinetic terms, and $s_V$ and $s_P$ represent the number of the total remaining terms after the potential and kinetic truncation, respectively. Here $n=8$, meaning 16 qubits are used in the quantum algorithm.

Therefore, a reasonable strategy is to keep all non-zero kinetic terms, which are already quite a few in number, and set an appropriate threshold for the potential terms to balance the error and the number of remaining terms. We also need to consider that the distributions of the coefficients vary with the qubit number.

Fig. 12 shows the truncation error and the total number of the remaining terms with different truncation thresholds when using 12, 14, 16 qubits. Overall, to maintain consistent relative error, as the qubit number increases, the threshold needs to be appropriately lowered to keep more terms. Here we provide an empirical formula by the three sample points closest to the dashed line in Fig. 12 to maintain approximately a 1% average relative error:

$$\tau_V = \frac{128}{2^n} \times 10^{-3}, \tag{17}$$

Of course, one can tweak this truncation threshold for a faster computation speed or a lower truncation error.

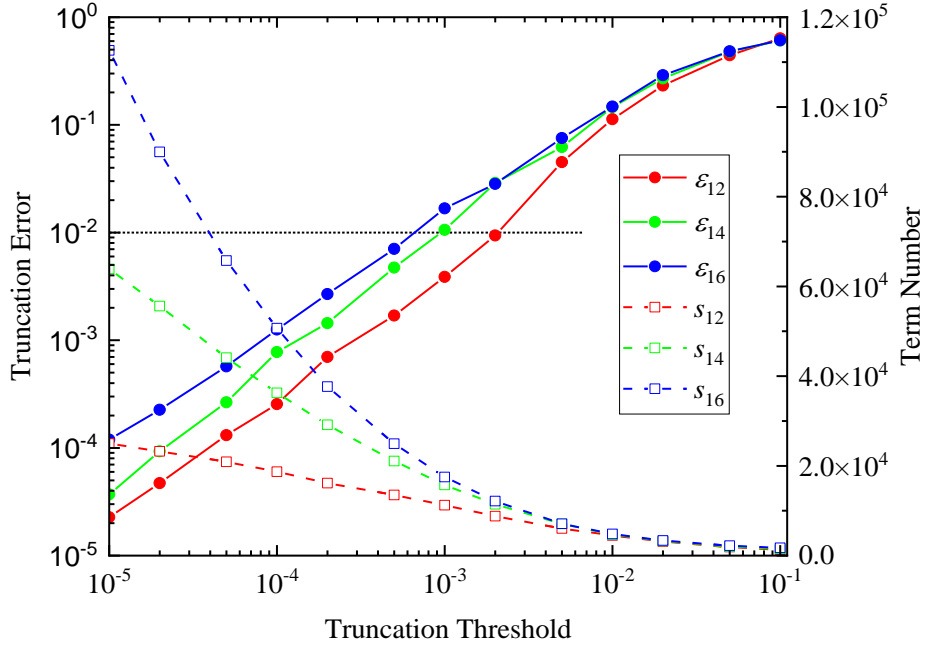

Fig. 12. The truncation error (denoted by $\varepsilon$) and the total term number (denoted by $s$) with different truncation thresholds for 12, 14, 16 qubit number. $\varepsilon_{12}$, $s_{12}$: 12 qubits, $n=6$; $\varepsilon_{14}$, $s_{14}$: 14 qubits, $n=7$; $\varepsilon_{16}$, $s_{16}$: 16 qubits, $n=8$.

Since as the number of qubits increases, we need to lower the truncation threshold to keep more potential terms, there is a possibility that, as the number of qubits continues to increase in larger simulations, the speedup of truncation may diminish. To examine this problem, we calculate the percentage of the remaining terms to the original terms at different qubit number, based on the empirical formula above to determine the truncation threshold. The results are shown in Fig. 13. As can be seen from the figure, although the truncation threshold is lowered and we keep more terms when the qubit number increases, the percentage of the remaining terms actually decreases. This indicates that the aforementioned problem does not exist; instead, the speedup effect of the truncation will improve as we use more qubits.

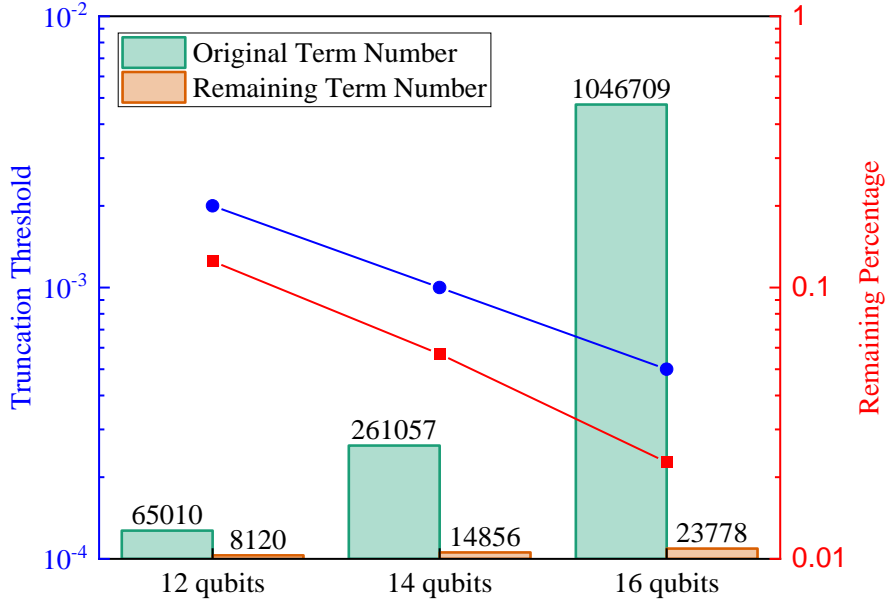

Fig. 13. The remaining percentage of terms and the truncation thresholds, ensuring that the truncation error is controlled around 1%.

We use this set of better truncation thresholds to plot the simulation results for different numbers of qubits and count the total number of quantum gates, comparing it with the situations without truncation optimization. The truncation threshold for the propagator, $\tau_P$, is set to $10^{-10}$, in order to keep all the non-zero terms. And the truncation threshold for the potential operator, $\tau_V$, is set according to Eq. (17). The simulation results are shown in Fig. 14. It can be seen that, with a relative error of about 1%, the approximate results are visually very close to the exact results, and keep consistent with different qubit number.

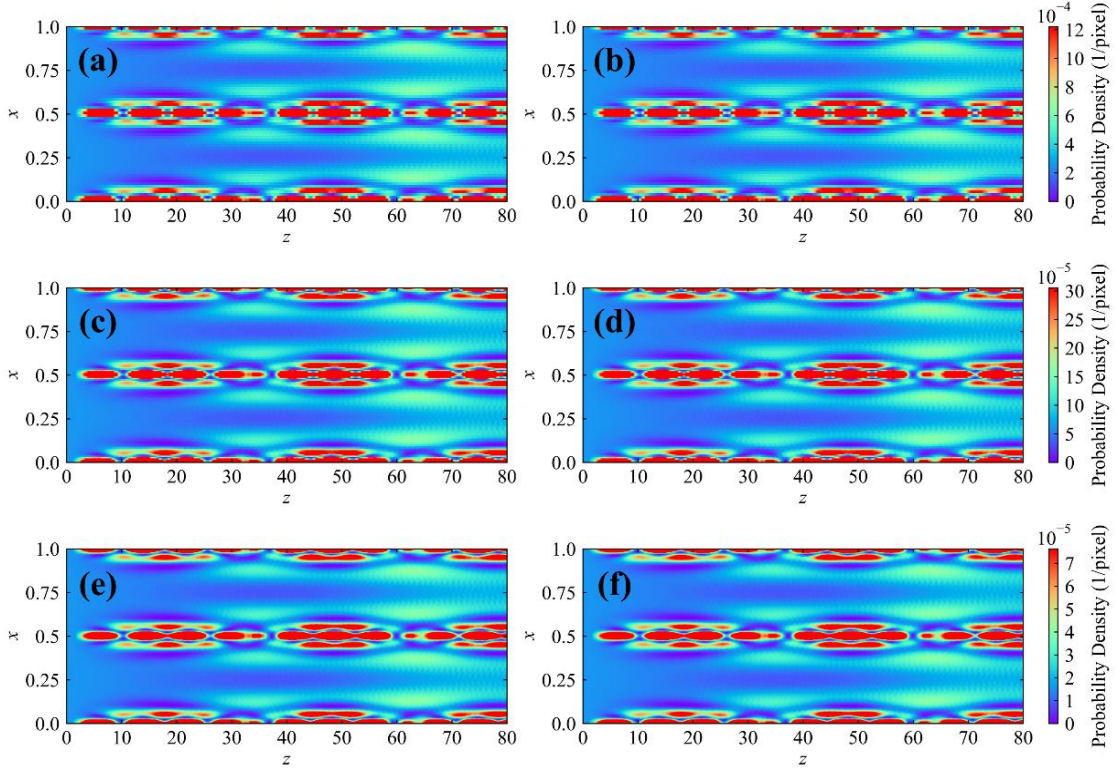

Fig. 14. (a) (c) (e) The exact simulation results without truncation and (b) (d) (f) the approximate simulation results using truncation optimization with different number of qubits: (a) (b) 12 qubits, (c) (d) 14 qubits, (e) (f) 16 qubits. $\tau_P$ is set to $10^{-10}$ in (b) (d) (f), $\tau_V$ is set to: (b) 0.002, (d) 0.001, (f) 0.0005.

We also count the number of quantum gates to quantify the computational cost of the quantum algorithm, as shown in Fig. 15. We can see that through truncation optimization, we have successfully reduced the number of quantum gates by more than one order of magnitude with controllably small error. Moreover, the slope of the line after truncation optimization is lower, indicating that the speedup effect may increase as the size of the quantum circuit grows.

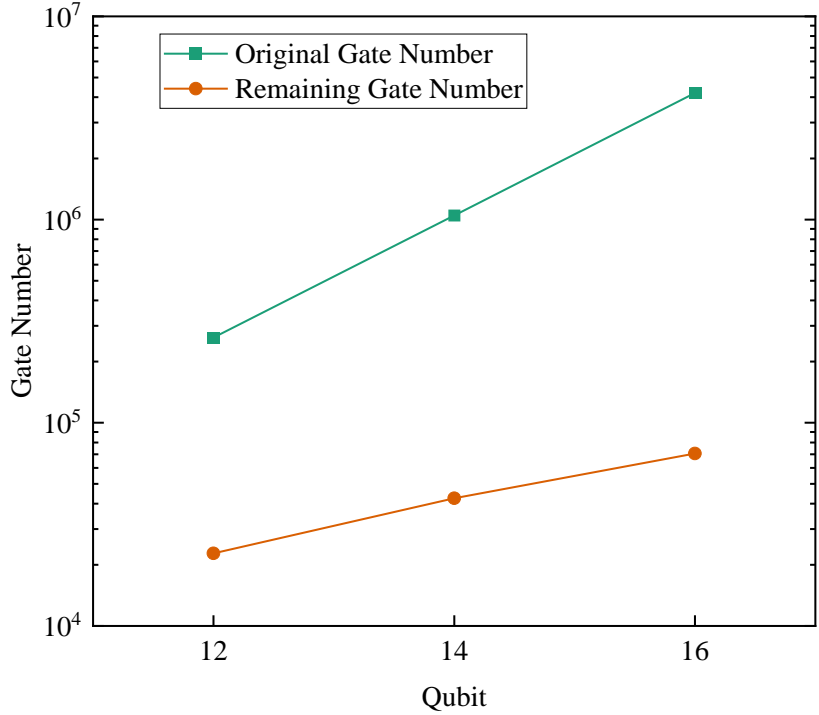

Fig. 15. Quantum gate numbers before and after truncation optimization.

## 5. Conclusion

In this paper, we present an improved quantum algorithm of the multislice method, which can be used to simulate the 3D scattering and diffraction of high-energy electrons with quantum circuits. This improved quantum algorithm is based on our previous version, with a focus on optimizing the phase-shifting circuit. Using the Walsh transform, we reconstructed the phase-shifting quantum circuit by replacing all multi-controlled gates with a comparable number of one-qubit and two-qubit gates, thereby addressing the potential performance issue of the previous quantum algorithm on practical quantum hardware.

The improved phase-shifting circuit also offers the possibility of further reducing the computational cost by truncating the Walsh terms with small coefficients. Through sufficient testing using different parameters, we have found a suitable truncation threshold setting scheme that can further reduce the number of quantum gates required by more than an order of magnitude while keeping the relative error on the order of 1%. And we show that this speedup effect is not weakened but enhanced as the number of qubits increases, ensuring that this truncation optimization can be used for larger-scale quantum simulations.

In our previous work, without considering the compilation issue of multi-controlled gates, we had initially achieved a complexity advantage relative to the classical multislice algorithm. And after the optimizations in this paper, the improved phase-shifting circuit, combined with the QFT which replaces the FFT, makes the overall complexity advantage of the improved quantum algorithm clearer. This demonstrates the potential application of quantum computing in the field of electron scattering and diffraction simulations and brings us closer to achieving quantum advantage using real quantum hardware in this area.

**Declaration of interests**

The authors declare no competing interests.

**Acknowledgement**

This work was supported by the National Natural Science Foundation of China (Grants Nos. T2422016, 42374108), the Fundamental Research Funds for the Central Universities (Grant No. 20720230014) and the Natural Science Foundation of Xiamen (Grant No. 3502Z202371007). We thank Prof. H.M. Li and the supercomputing center of USTC for the support of parallel computing.